\def\versiondate{{3 April 1997}}
\def\titleabbrev{{Gravitational Thermodynamics}}
%
%
\input psfig
%
%
\def\undertext#1{{$\underline{\hbox{#1}}$}}
\def\doubleundertext#1{{$\underline{\underline{\hbox{#1}}}$}}

\newcount\notenumber
\def\clearntnumber{\notenumber=0}
\def\nt{\advance\notenumber by1 \footnote{$^{\the\notenumber}$}}
\clearntnumber

\def\simlt{\hbox{ \rlap{\raise 0.425ex\hbox{$<$}}\lower 0.65ex
  \hbox{$\sim$} }}
\def\simgt{\hbox{ \rlap{\raise 0.425ex\hbox{$>$}}\lower 0.65ex
  \hbox{$\sim$} }}

%
%
\magnification=1200
\hsize=4.5in
\hoffset=0.5in
\vsize=6.9in
\voffset=0.3in

\font\lggb=cmbx10 scaled \magstep3

\font\lb=cmbx10 scaled \magstep1

\font\lr=cmr10 scaled \magstep1
\font\slgh=cmr10 scaled \magstep2
\font\sb=cmbx9
\font\sr=cmr9
\def\part#1{\vfill\eject\hbox{}\bigskip\bigskip
  {\parindent=0pt\lb #1}\bigskip}
\def\chapter#1{\hbox{}\bigskip\goodbreak
  {\parindent=0pt\bf #1}}
\def\beginchapter#1{
  {\parindent=0pt\bf #1}}

\def\notestart{{\parindent=0pt\medskip\item{\undertext{\sb Notes}}
  {}\par}}
\def\note#1#2{{\parindent=0pt\medskip\item{\sr [#1]}{\sr #2}\par}}
\def\ref{{\parindent=0pt\medskip\hangindent=3pc\hangafter=1}}
\hyphenation{Des-car-tes}
\hyphenation{phe-no-me-non}
\baselineskip 13pt
\parskip 5pt plus 1pt
    %
    %
\hbox{}
\bigskip
\bigskip
\bigskip
\bigskip
\centerline{\lggb
    Gravitational Thermodynamics     
}
\bigskip
\bigskip
\bigskip
\bigskip
\bigskip
\centerline{\slgh Piet Hut}
\bigskip
\bigskip
\centerline{\lr Institute for Advanced Study}
\smallskip
\centerline{\lr Princeton, NJ 08540, U.S.A.}
\bigskip
\centerline{{\it email:} piet@ias.edu}
\bigskip
\def\boxit#1{\vbox{\hrule\hbox{\vrule\kern3pt
             \vbox{\kern3pt#1\kern3pt}\kern3pt\vrule}\hrule}}
\setbox4=\vbox{\hsize 5pc \noindent \strut
To appear in: \hbox{\ \ {\it Complexity}}
\strut}
$$\boxit{\boxit{\box4}}$$
\bigskip
\bigskip
\bigskip


\chapter{Abstract}

     The gravitational $N$-body problem, for $N>2$, is the
oldest unsolved problem in mathematical physics.  Some of the
most ideal examples that can be found in nature are globular star
clusters, with $N \sim 10^6$.  In this overview, I discuss six
types of fundamental sources of unpredictability, each of which
poses a different challenge to attempts to determine the
long-term behavior of these systems, governed by a peculiar
type of thermodynamics.

\nopagenumbers
\vfill\eject
\headline={\tenrm \doubleundertext\titleabbrev \hss
   \undertext\versiondate} 
\footline={\hss \tenrm \folio \hss}

    \beginchapter{1. Introduction}

     The theory of the gravitational dynamics of celestial
bodies was first put on a firm footing by Newton in 1687, and
was subsequently applied, both by him and by many others, to
phenomena in the solar system.  These particular applications,
that form the discipline of celestial mechanics, were very
specialized.  Planets and asteroids move in orbits that
mostly lie near a common plane, and their mutual perturbations
are small compared to the dominant gravitational attraction of
the much more massive Sun.  And while comets describe orbits
with a higher eccentricity and inclination, their masses are so
small that they are effectively test particles, reacting to the
force field around them but exerting no appreciable forces on
their environment.

     It took two and a half centuries before gravitational
dynamics grew up, out of this cradle of celestial mechanics,
into a full-fledged theory of stellar dynamics, able to
describe self-gravitating configurations of particles without
any a-priori restrictions.  Indeed, a typical star cluster
consists of a group of stars without any central dominant body
such as the sun, and without any preferred plane of motion,
such as that which characterizes the motion of the planets.  A
stellar dynamical description of a star cluster thus needs to
be statistical in nature, and as such, it had to await the
advent of thermodynamics, two centuries after Newton.

     As we will see, the greater simplicity of stellar
interactions over molecular interactions paradoxically created
extra difficulties in attempts to apply laboratory thermodynamics
to self-gravitating star systems.  As a result, it was only
toward the middle of this century that significant progress was
made on a theoretical level.  Actual experimentation had to
wait even longer, until computers were powerful enough.  Some
of the most provocative theoretical ideas are just now becoming
amenable to thorough testing in large-scale simulations of the
full gravitational many-body problem, with particle numbers in
the range 10,000 to 100,000.\nt{These numbers are the state of
the art for simulations where each particle interacts with each
other particle, for many relaxation times.  Much more specialized
calculations, such as those modeling structure formation in the
early universe, can make use of various approximations that
significantly reduce computer time, allowing total particle
numbers up to $10^7$.}

    \chapter{2. Fundamental Sources of Uncertainty}

     The gravitational many-body problem addresses the behavior
of an isolated system of point masses held together by
Newtonian gravitational forces.  There are several fundamental
sources of uncertainty that stand in the way of a detailed
description of the evolution of such a system.  Three of those
sources are directly related to predictability, while three
others exist that limit an effective thermodynamic treatment of
a self-gravitating system of point particles.

     On a microscopic level, neighboring orbits show an
exponential divergence with a Lyapunov time scale less than the
time it takes for a typical particle to cross the system.  In
addition, bound pairs of particles, when encountering other
particles or particle pairs, typically engage in complex
interactions that exhibit an extreme sensitivity to initial
conditions.  On a macroscopic level, chaotic core oscillations
in the dense center of the system introduce extra degrees of
uncertainty into our attempts to model the detailed evolution
of the system.

     One might have hoped that these three sources of
uncertainty would still admit an average overall understanding
of the system in the form of a thermodynamic treatment.
However, there are three other obstacles that stand in the way
of a full-fledged theory of gravitational thermodynamics.  For
starters, we cannot define a thermodynamic limit, since
gravitational binding energy grows faster than the mass of the
system, when we increase the scale (in particle physics
terminology, an infrared divergence).  Furthermore, the system
can exhibit an effectively negative heat capacity, which is
related to the fact that a close pair of particles can generate
an arbitrary amount of energy (an ultraviolet divergence).
Finally, a self-gravitating system has only one coupling
constant, which does not allow decoupling and fine-tuning of
microscopic and macroscopic behavior separately.  As a result,
we are left with no degrees of freedom to play with.  There are
no dials that one can turn in order to study the response of
the system.

     In this overview, in \S\S3-8 each of these six
characteristics will be discussed briefly in order to give some
flavor of the difficulties involved.  Notwithstanding these
difficulties, gravitational many-body systems are realized in
nature, and can indeed be modeled to high accuracy in computer
simulations.  These will be briefly reviewed in \S\S9,10.

    \chapter{3. Exponential Instability}

     In order to explore the behavior of a self-gravitating
system of point masses, let us start with a thought experiment.
Let us call our point particles `stars', in order to anticipate
astrophysical applications, even though at this point we
neglect any physical properties the particles may have other
than their mass.

     Take a number of stars, a million say (a modest number,
given that a typical galaxy contains $10^{11}$ stars).  If we
throw them into empty space, far away from disturbing tidal
forces due to neighboring galaxies, the initial clump of stars
may well look quite irregular at first.  Since this arbitrary
configuration is not likely to be near an equilibrium state,
the whole system will undergo a few oscillations, and perhaps
expand or contract considerably, depending on the initial
kinetic energy given to the particles.  However, after a few
crossing times (the time $t_{cr}$ it takes for a typical star
to cross the bulk of the star system) these oscillations will
be damped out, largely through phase mixing [1]

     From this point onwards, the resulting star cluster will
look quite smooth.  It may still be flattened or elongated
considerably, and it may exhibit some bulk rotation.
Subsequent two-body relaxation, exchange of energy and angular
momentum between pairs of stars passing each other, tends to
erase any such memory of the initial conditions.  These effects
take place on a two-body relaxation time scale $t_{rel}$, which
can be shown to be related to the crossing time $t_{cr}$ as
$$
t_{rel} \sim 0.1 {N\over\ln N} t_{cr}.
$$
For a system of a million particles, the two time scales are
thus well-separated, by some four orders of magnitude.

     Analytical estimates, as well as more detailed computer
simulations, can predict the overall qualitative and
semi-quantitative effects of the process of two-body
relaxation.  However, a really accurate modeling is not
feasible, even on the fastest computers.  The problem is that
the slightest change in the initial position or velocity of a
single particle leads to an exponentially growing divergence of
the trajectories of all particles, compared to the evolution of
the original system.  The e-folding time scale $t_e$ for this
divergence was recently determined [2] to be
$$
t_e \sim 0.2 {1\over\ln\ln N} t_{cr},
$$
nearly, but not quite, independent of the total number of stars.

     The evolution of a star cluster takes place on a time
scale far longer than $t_e$.  Heat is transported through the
cluster, as a consequence of many two-body encounters, on the
time scale $t_{rel}$.  On longer time scales, any
self-gravitating star system is unstable.  While some stars in
the Maxwellian tail of the velocity distribution are lost
through escape, other stars tend to congregate in the central
regions which grow denser at an ever-increasing rate, since
higher density implies more frequent encounters and hence a
faster two-body relaxation.

     This run-away redistribution of energy and mass leads to a
phenomenon called gravothermal collapse, often called core
collapse for short, which takes place on a time scale $t_{cc}
\sim 10 t_{rel}$.  Predicted analytically in the sixties, seen
in approximate numerical simulations in the seventies, and
verified in direct $N$-body simulations in the eighties, core
collapse is a fundamental feature of long-term
stellar-dynamical evolution.  Its occurrence is related to the
fact that series of self-gravitating equilibrium models exhibit
a maximum entropy for a finite central concentration.  There is
no room here to go very deeply into the fascinating physics
behind this phenomenon.  For a general description, and
references to the literature, Spitzer's concise monograph on
globular clusters would be a good starting point [3].

     Suffice to say here that simple estimates, based on
two-body encounters alone, would predict an infinitely high
central density to develop after a time $t_{cc}$.  In physics,
any prediction giving infinite numbers suggests that some
additional physics is needed.  In this case simultaneous
three-body encounters can save the situation, as we will see in
the next section.

     How serious is the uncertainty introduced by the
exponential divergence of neighboring trajectories?  If we want
to simulate the evolution of a star cluster on a computer, all
the way to core collapse and beyond, we have to overcome a
number of e-folding times of order
$$
{t_{cc} \over t_e} \sim 5N {\ln \ln N \over \ln N} \sim N,
$$
where the last approximate equality holds to within a factor
two for particle numbers throughout the astronomically
interesting range $10^2 \sim 10^{12}$.  Since we lose roughly
one digit of accuracy for every two e-folding times, a
simulation of core-collapse of a self-gravitating $N$-body
system thus requires us to calculate each orbit to a fantastic
accuracy.  Rather than working in the usual double-precision
mode, where each number is given to about 15 digits, we need
a word-length of $\sim N/2$.  Even if we use a 128-bit word
length, we can only accurately model systems with up to 60
particles.  On the fastest computers available, a simulation to
core collapse thus turns out to be totally infeasible for any
realistic value of $N > 10^3$, let alone $N=10^6$ as is
appropriate for globular star clusters, the primary examples of
isolated star systems.

    \chapter{4. Three-Body Effects}

     Even in an idealized system of self-gravitating point
particles, core collapse will be halted before an infinite
central density is reached.  When the central density is high
enough, occasional close encounters between three unrelated
particles will form bound pairs (double stars in the case of
star clusters), with the third particle carrying off the excess
kinetic energy required to leave the other two particles bound.
Subsequent encounters between such pairs and other single
particles tend to increase the binding energy of these pairs,
which leads to a heating of the surrounding system of single
particles.

     When enough pairs have been formed this way, the resulting
energy production will reverse the process of core collapse.
After reaching a minimum radius and maximum density, the core
region will expand again.  An accurate modeling of this process
of deep collapse and subsequent re-expansion is even more
taxing than a collapse simulation.  Besides the exponentially
unstable trajectories of single particles, we now have to model
the encounters between singles and pairs, as well as those
between two pairs and occasionally even more complicated
interactions.

     Most frequent are the three-body interactions resulting
from a pair-single encounter.  In many of these cases, the
third body is caught into a bound orbit, and the whole group of
three begins an erratic gravitational dance stretching out over
an extended period, before one of the three particles is
ejected.  During this dance, close encounters between two of
the three particles often number in the tens or hundreds, if
not thousands, and any small deviation in the initial
conditions will result in a vastly different outcome.  Typical
amplification factors are $10^{10} \sim 10^{20}$, and factors
exceeding $10^{100}$ are not uncommon [4].  A faithful
simulation of these temperamental situations would require far
larger word lengths than the standard 64-bit (15-digit) `double
precision' word length implemented in most computers.

    \chapter{5. Gravothermal Oscillations}

     Core collapse, when threatened to occur by the collective
effects of two-body relaxation, can thus be narrowly averted by
a handful of crucial three-body or four-body reactions in the
very dense core of a near-collapsed cluster.  What will happen
next depends on the total number $N$ of particles in the
system.  If this number is sufficiently small, $N < 10,000$ or
thereabouts, the whole system will slowly and steadily expand.
In this case an equilibrium can be found between the steady
energy production in three-body encounters in the center, and
the continuous loss of energy through the outskirts of the
system.

     If the total number of particles somewhat exceeds $10^4$,
however, a different behavior emerges.  The more particles
there are in the system, the higher the central density has to
become in deep core collapse, to be able to hold the initial
contraction.  As a result, the post-collapse phase features a
very short central relaxation time, far shorter than the
relaxation time in the outer regions, where most of the
particles can be found.  From the point of view of the inner
dynamics, the bulk of the mass further out seems almost frozen.
It is this discrepancy in time scales that can cause the inner
core to become `impatient', and to revert to a local collapse,
triggered by the slightest fluctuation in the direction of the
energy flow produced by stochastic three- and four-body
interactions.

     What happens then is that the inner one percent or so of
the total number of particles will go once more into a coherent
collapse, locally reminiscent of the original deep
collapse.\nt{Note that here and elsewhere the established
jargon is somewhat misleading: the so-called `collapse' occurs
on a local relaxation time scale which even in the center is
far larger than the local crossing time, except around core
bounce -- little happens during the time in which a typical
particle traverses the system, since it is the thermal
equilibrium that is perturbed, not the dynamical equilibrium.}
As before, bound pairs of particles spring into action,
generate energy, and manage to reverse the collapse in the nick
of time, preventing an infinite central density from building
up.  This process repeats itself, leading to irregular
oscillations.  The chaotic nature of these bulk oscillations
forms a third source of fundamental uncertainty, preventing
detailed predictions to be made of the behavior of a
self-gravitating system of point masses, even on a macroscopic
level.

     Although we are dealing here with the oldest unsolved
problem in mathematical physics, the behavior of a system of
$N$ gravitationally interacting particles for $N>2$, the
existence of these oscillations was completely unknown until
1983, when they were first found in approximate simulations
[5].  Dubbed `gravothermal oscillations', they were
subsequently analyzed in detail with semi-analytic methods [6].
Their occurrence was confirmed in a variety of approximate
numerical simulations [7], and shown to correspond to
low-dimensional chaos for large $N$ values [8].  Finally, their
existence was proven beyond the shadow of a doubt when they
were seen in the very first direct $N$-body simulation that was
able to incorporate $N$ values well beyond $N=10,000$ [9].
Interestingly, it took the construction of a special-purpose
computer, with a peak speed of 1 Teraflops, to pull off this
feat [10].

    \chapter{6. No Thermodynamic Limit}

     Quite apart from the three problems raised above, there
are more fundamental problems that prevent a full thermodynamic
treatment of a self-gravitating system of point particles.
First of all, gravity exhibits what is known in particle
physics as an infrared divergence.  This means that the effect
of long-distance interactions cannot be neglected, even though
gravitational forces fall off with the inverse square of the
distance.

     Take a large box containing a homogeneous swarm of stars.
Now enlarge the box, keeping the density and temperature of the
star distribution constant.  The total mass $M$ of the stars
will then scale with the size $R$ of the box as $M \propto
R^3$, and the total kinetic energy $E_{kin}$ will simply scale
with the mass: $E_{kin} \propto M$.  The total potential energy
$E_{pot}$, however, will grow faster: $E_{pot} \propto M^2/R
\propto M^{5/3}$.  Unlike intensive thermodynamic variables
that stay constant when we enlarge the system, and unlike
extensive variables that grow linearly with the mass of the
system, $E_{pot}$ is a superextensive variable, growing faster
than linear.

     As a consequence, the specific gravitational potential
energy of the system, the total potential energy of the system
divided by the particle number $N$, grows without bounds when
we increase $N$.  This causes various problems.  For example,
the kinetic energy of a stable self-gravitating system is
directly related to the gravitational potential energy through
the so-called virial theorem.  Therefore, we have to make a
choice when enlarging a self-gravitating system.  Either we
increase the temperature steadily while increasing $N$, in
order to increase the specific kinetic energy enough to satisfy
the virial theorem and guarantee stability.  Or we keep the
temperature constant, and quickly lose stability when enlarging
our system.  In the latter case, the system will `curdle': it
will fall apart in more and more subclumps, and the original
homogeneity will be lost quickly.

     In conclusion, there is no way that we can reach an
asymptotic thermodynamic limit, which the system size becoming
arbitrarily large while holding the intensive variables fixed.
Therefore, the traditional road to equilibrium thermodynamics
is blocked.  There are no arbitrarily large homogeneous
distributions of stars.  As soon as the Universe became old and
cold enough to let matter condense out of the original fire
ball into `islands' in the form of galaxy clusters and
galaxies, the original homogeneity was lost.  And each
individual clump of self-gravitating material, be it a galaxy
or a star cluster, is ultimately unstable against evaporation,
and will fall apart into a bunch of escaping particles.  Most
of these escapers will be single, some will escape as stable
pairs, and a few will even manage to form stable triples or
higher-number multiples of particles.

    \chapter{7. Negative Heat Capacity}

     Another problem interfering with a standard thermodynamic
treatment of the gravitational many-body problem is what is
known in particle physics as an ultraviolet divergence.  Since
the individual particles are mass points with no spatial
extension, they can come arbitrarily close, and therefore their
negative gravitational binding energy can become arbitrarily
large.  Just one pair of particles can therefore provide an
unlimited amount of positive energy to the rest of the system.

     This is dramatically illustrated in the following thought
experiment.  Let us confine $N$ point masses to a box with
reflecting walls, while we neglect the gravitational effects
due to the finite mass of the walls.  If we wait long enough, a
simultaneous close three-body encounter will produce a tightly
bound pair.  It can be shown that from that moment on, the
probability is overwhelmingly large that this pair will grow
tighter and tighter, on average, giving off more and more
energy.  This energy is converted to kinetic energy of all the
other particles, including the kinetic energy of the
center-of-mass motion of the bound pair.

     In this thought experiment, after closing the lid of the
box, we would notice that the walls of the box would get
hotter, without bound.  Even if we would slowly extract heat
from the box, its temperature would keep rising.  The cause of
this paradoxical behavior is due to the fact that a
self-gravitating system can effectively have a negative heat
capacity.  The tightly bound particle pair in our experiment is
one example of such a system.  Its orbital motion will be much
faster than the movement of the single particles that it
encounters.  In an attempt to reach equipartition, the bound
particles will try to convey some of their rapid motion to the
single particles, speeding the latter up in the process.  The
bound particles themselves, however, while attempting to slow
down, will find themselves falling to an even tighter orbit.  A
shorter distance in the gravitational two-body problem implies
a higher orbital velocity, so the net effect is that {\it both}
the single particles and the bound particles will speed up as a
result of their interactions.

    \chapter{8. A Single Coupling Constant}

     The motion of the stars in a star cluster can be described
in a way that is analogous to the treatment of the motion of
molecules in a gas studied in a laboratory.  One important
difference is that a swarm of stars forms an open system, while
a body of gas in a lab has to be contained.  Typical textbook
experiments in thermodynamics show the gas to reside inside a
cylinder, with a movable piston that allows the experimenter to
change the volume of the gas.  In a star cluster, there are no
cylinder and piston.  Instead, the stars are confined by their
collective gravitational field.

     The structural simplicity of a star cluster thus allows
far less experimentation than is the case for a body of gas in
a lab situation.  Whether in thought experiments, computer
simulations, or in actual table top experiments, the
macroscopic parameters of a laboratory gas can be changed
freely, independent of the microscopic parameters governing the
attraction and repulsion between individual molecules.
Temperature, density, and size of the system all can be varied
at will.  In contrast, once the number of particles in a
self-gravitating system has been chosen, we are left with no
degree of freedom at all, apart from trivial scalings in the
choice of unit of length, time, and mass.

     The fact that there are no dials that can be turned in a
self-gravitating experiment, apart from the choice of the total
number of stars, is directly related to the ultraviolet and
infrared divergences of classical gravity.  Having a simple
shape for the gravitational potential energy well, with an
energy inversely proportional to distance, leaves no room for
preferred length scales.  In contrast, molecular interactions
show far more complicated forces, typically strongly repulsive
at shorter distances and weakly attractive at larger
separations between the molecules.  This change in behavior
automatically specifies particular length scales, for example
the distance at which repulsion changes into attraction.  In
contrast, gravity is attractive everywhere, at least in the
classical Newtonian approximation.

    \chapter{9. Astrophysical Applications}

     Gravitation is a purely attractive force.  Even though
electromagnetism is far stronger than gravity, on microscopic
length scales as well as on normal human scales,
electromagnetic effects tend to cancel out on much larger
distances.  As a result, gravitation plays a dominant role in
almost all astrophysical objects, from planets to stars to star
clusters to galaxies and clusters of galaxies.

     Not all of these objects can be described accurately with
the simple models that we have discussed so far.  Galaxies, for
example, contain massive gas clouds whose hydrodynamic
interactions are as important as their gravitational
properties.  Furthermore, the individual galaxies in a cluster
of galaxies are significantly extended, in comparison with the
intergalactic distances, and therefore cannot be treated very
well as point particles.

     The ideal type of astrophysical objects, from the point of
view of studying gravitational thermodynamics, are the globular
clusters (fig. 1).  Each one contains of order a million stars.
Other than that, they contain very little gas, are far removed
from the bulk of the stars in our galaxy, and are among the
oldest objects in the Universe.  All these properties make them
ideal laboratories for stellar dynamics [3].  There are about a
hundred of these clusters circling our own galaxy.  While many
other galaxies are accompanied by similar numbers of globulars,
some galaxies are adorned by thousands or more of them.

\bigskip
\bigskip
\centerline{\psfig{figure=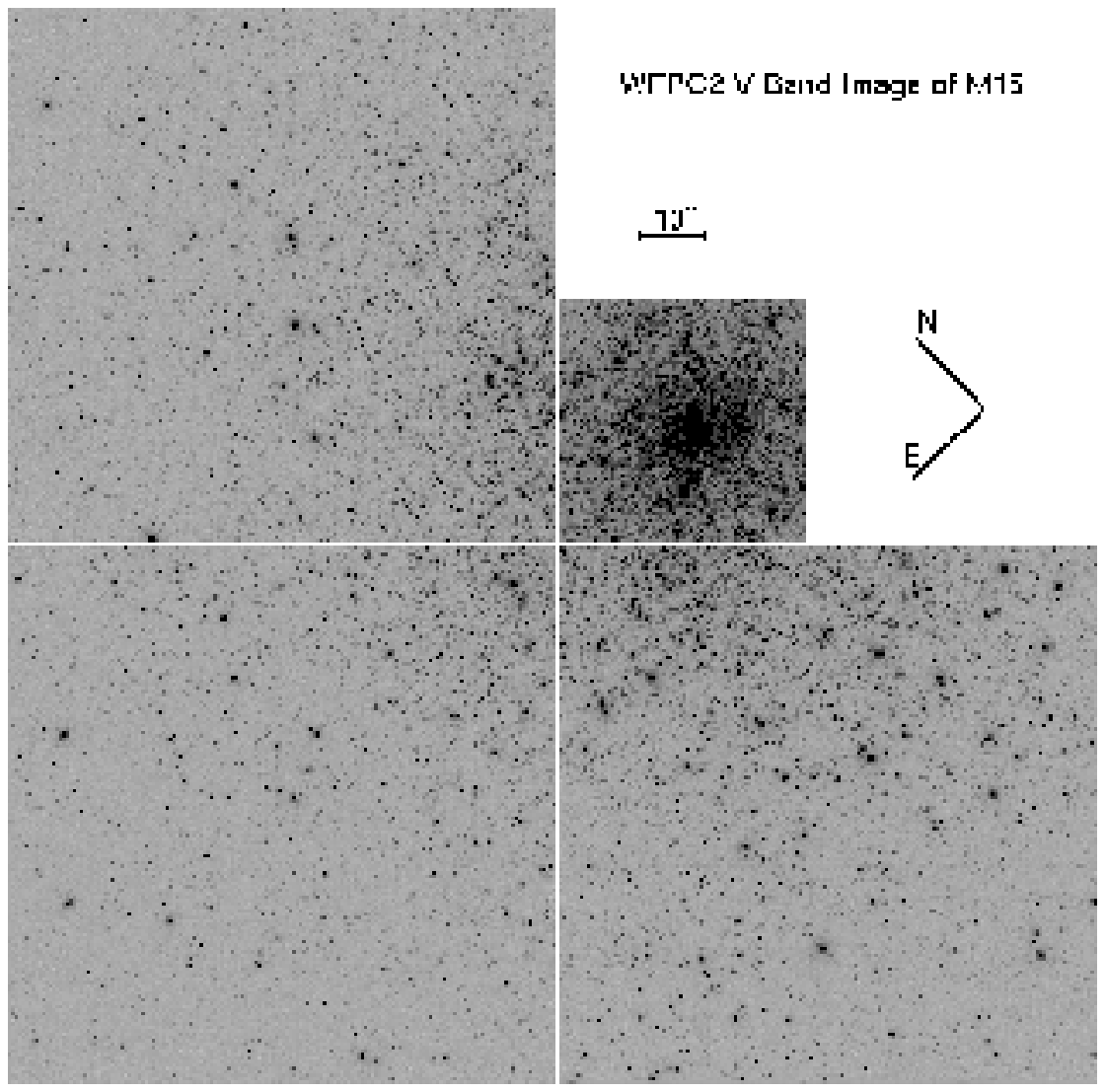,width=4.5in}}

\noindent{Fig. 1. M15, a typical globular cluster with a
collapsed core, as seen by the Hubble Space Telescope (in the V
band)[11].}

     Recently, the Hubble Space Telescope has allowed us a few
long-awaited peeks into the very densest inner regions of some
of the core-collapsed globular clusters (fig. 2).  Fortunately,
we are ready to interpret this wealth of new data.  As
discussed above, we have made great progress in our theoretical
understanding of the long-term evolution of self-gravitating
systems during the eighties.  In addition, computers have now
become powerful enough to enable realistic simulations to be
carried out.

\bigskip
\bigskip
\centerline{\psfig{figure=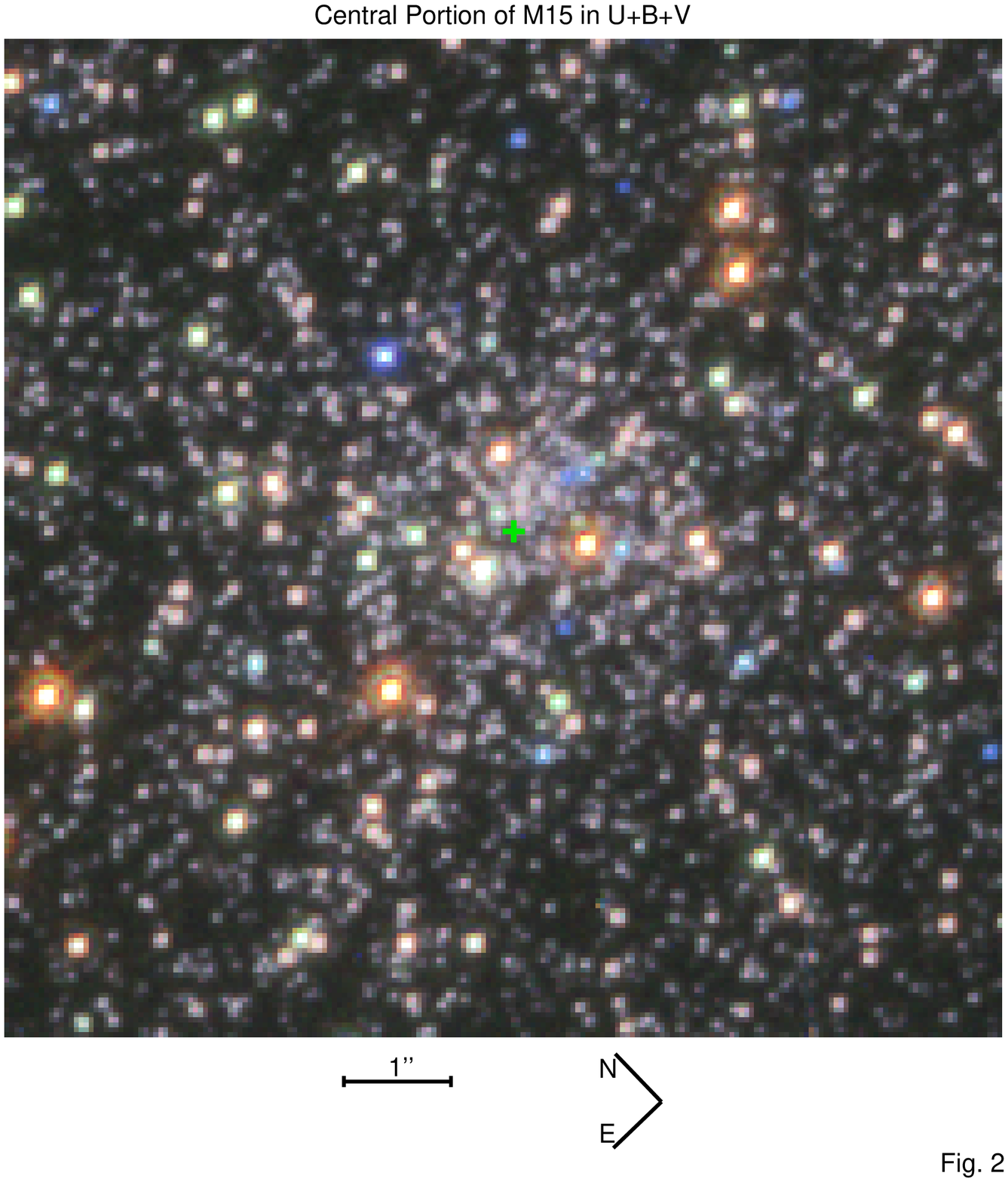,width=4.5in}}

\noindent{Fig. 2. A high-resolution view of the dense central
regions of M15; as in the previous figure, but with three
colors (U, B, and V), covering only the central 9" by 9".[11]}

    \chapter{10. Computer Simulations}

     In the late eighties, detailed analysis made it clear that
realistic simulations of globular clusters on a star by star
basis, necessary to investigate gravothermal oscillations,
would remain beyond the power of even the most advanced
computers for at least a decade [12].  Fortunately, a small
group of astrophysicists at Tokyo University decided to build
their own specialized computer for stellar dynamical
calculations.  Recently, their completed product (fig. 3), the
GRAPE-4 [10], reached a speed of 1 Teraflops, which made it the
fastest computer in the world.

\bigskip
\bigskip
\centerline{\psfig{figure=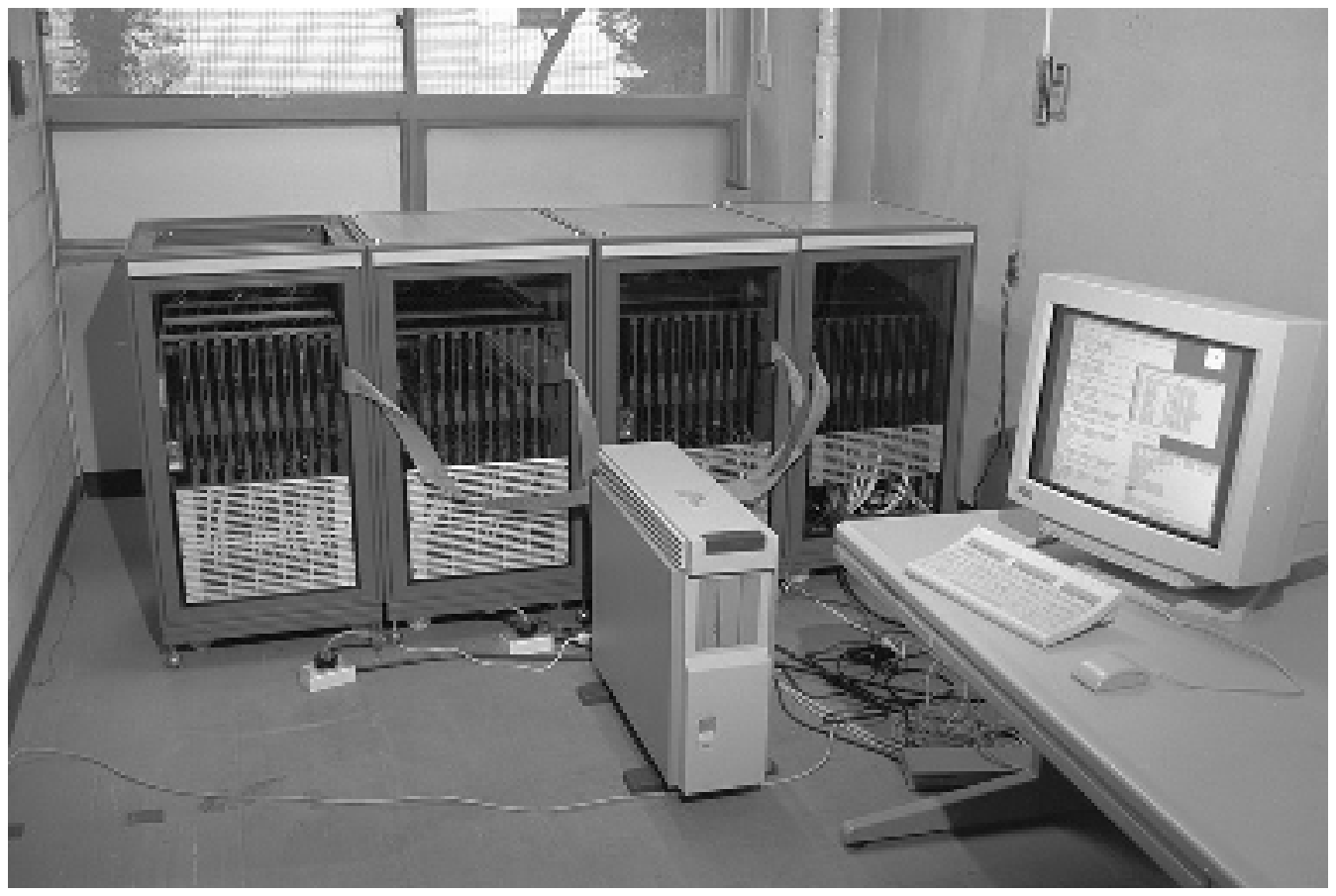,width=4.5in}}

\noindent{Fig. 3. The GRAPE-4, a special-purpose computer for
stellar dynamics simulations, operating at a speed of 1
Teraflops, record holder of the fastest computations in
1995 and 1996, for which two Gordon Bell prizes were
awarded to Junichiro Makino, seen here in the photograph,
together with Makoto Taiji (in 1995) and Toshiyuki Fukushige
(in 1996).}

     This computer allowed an unambiguous demonstration of the
reality of gravothermal oscillations (figs. 4,5).  The next
generation special-purpose computer, the GRAPE-6, is planned to
have a speed which is far higher than the GAPE-4, in the range
of 100-1000 Teraflops.  This will enable realistic simulations
of up to a million point particles.  If funding can be found,
the GRAPE-6 could be built within a few years.

\bigskip
\bigskip
\centerline{\psfig{figure=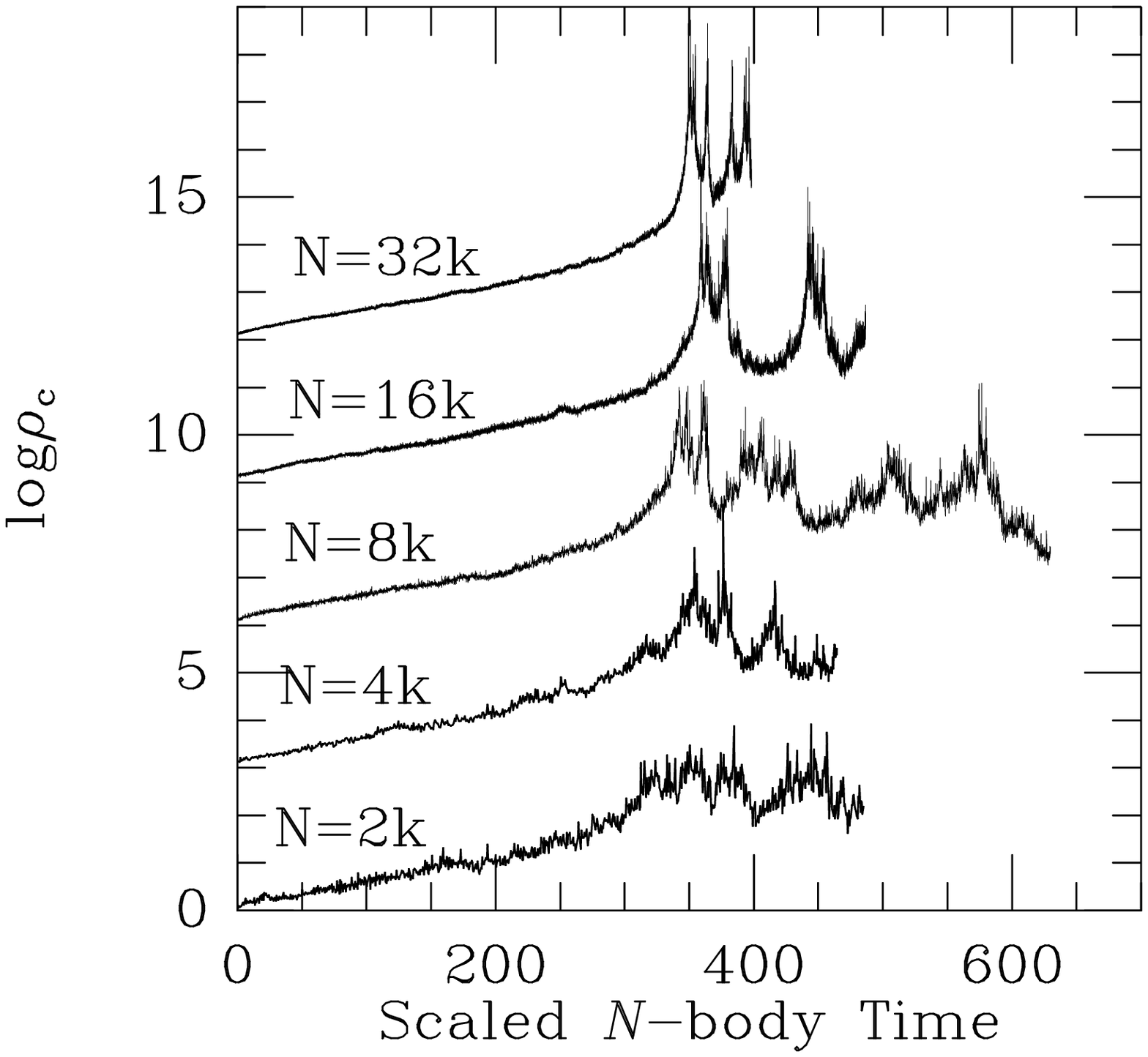,width=4.5in}}

\noindent{Fig. 4. The evolution of the central density in a
32,000-body problem, the result of a Teraflops-week calculation
on the GRAPE-4.  For comparison, calculations for smaller
number of particles are shown as well, in which case the
peak densities reached are lower.}

\bigskip
\bigskip
\centerline{\psfig{figure=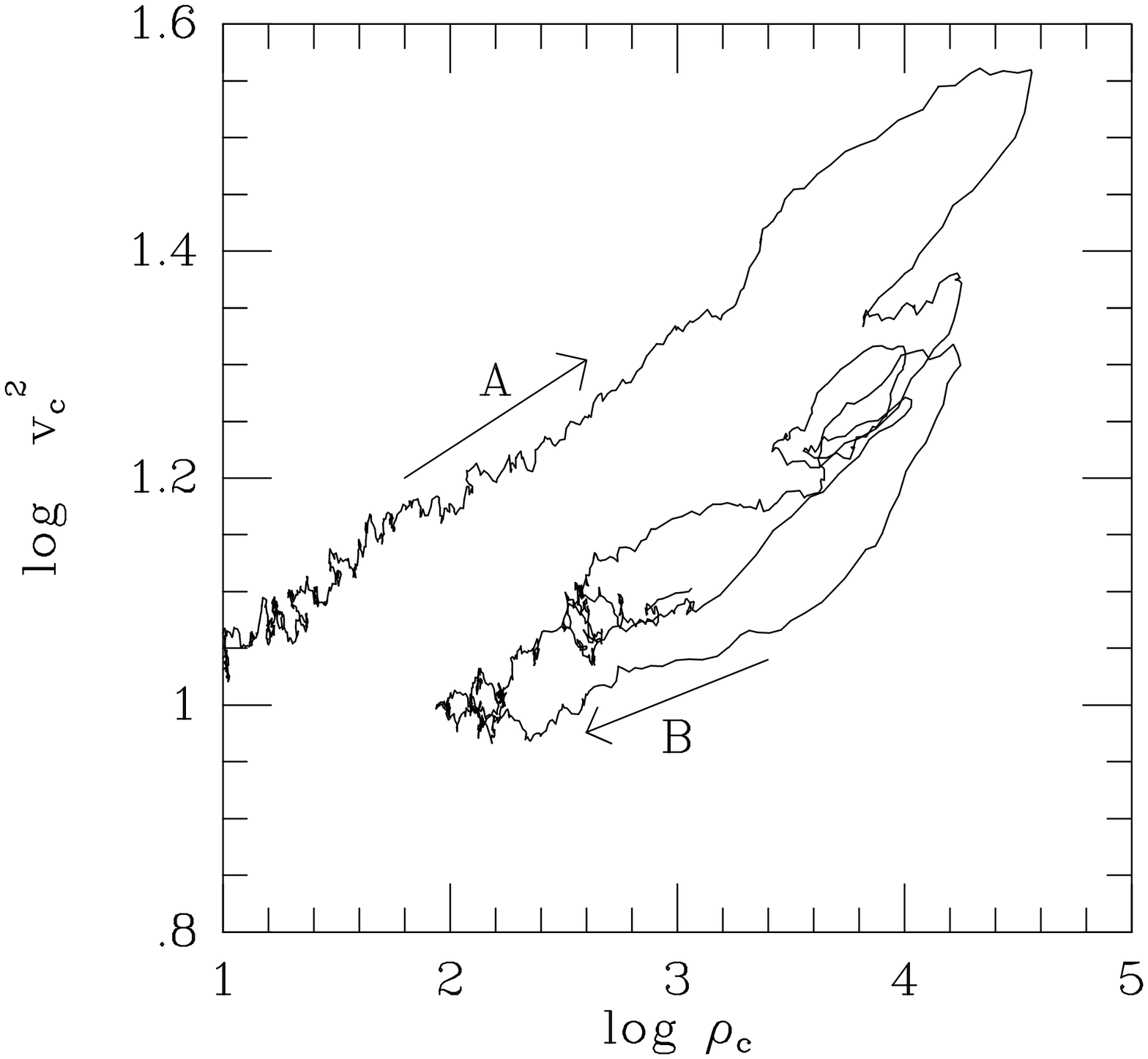,width=4.5in}}

\noindent{Fig. 5. The evolution of central density $\rho_c$ and
temperature (proportional to $v_c^2$, where $v_c$ is the
velocity dispersion), for the 32,000-body calculation of
Fig. 4, showing the first direct evidence for the existence of
gravothermal oscillations.}

     While point-particle simulations are of great interest
from the point of view of mathematical physics, astrophysics
requires a more realistic treatment.  Real stars have finite
radii, and therefore can run into each other and merge.  Also,
they age and die, ending their life either in a violent
supernova explosion, which may produce a neutron star or black
hole or lead to a total annihilation of the star, or by
settling down into a more quiescent final state as a white
dwarf.  These more realistic simulations are currently
underway [13,14].

    \chapter{11. Conclusions}

     The fact that a self-gravitating system of point masses is
governed by a law with only one coupling constant has important
consequences.  In contrast to most macroscopic systems, there
is no decoupling of scales.  We do not have at our disposal
separate dials that can be set in order to study the behavior
of local and global aspects separately.  The only real freedom
we have, when modeling a self-gravitating system of point
masses, is our choice of the value of the dimensionless number
$N$, the number of particles in the system.

     The value of $N$ determines a large number of seemingly
independent characteristics of the system: its granularity and
thereby its speed of internal heat transport and evolution; the
size of the central region of highest density after the system
settles down in an asymptotic state; the nature of the
oscillations that may occur in this central region; and to a
surprisingly weak extent the rate of exponential divergence of
nearby trajectories in the system.  Independent of particle
number, however, any self-gravitating system in or near
equilibrium exhibits a most unusual thermodynamic
characteristic, in the form of a negative heat capacity.

\bigskip
\bigskip
\noindent
{\sl Acknowledgments.}
I thank Douglas Heggie for comments on the manu\-script.
This work was supported in part by a grant from the Alfred
P. Sloan foundation, for research on limits to scientific
knowledge.  I thank the Santa Fe Institute for their
hospitality during my visits in March and June 1996.


\bigskip
\bigskip

\notestart

\note{1}{J. Binney \& S. Tremaine, Galactic Dynamics (Princeton:
Princeton Univ. Press), 1987.}

\note{2}{J. Goodman, D. C. Heggie \& P. Hut, On the Exponential
Instability of N-Body Systems, Astrophys. J., 415: pp. 715-733,
1993.}

\note{3}{L. Spitzer, Dynamical Evolution of Globular Clusters
(Princeton: Princeton University Press), 1987.}

\note{4}{H. Dejonghe \& P. Hut, Round-off Sensitivity in the
$N$-Body Problem, in {\it The Use of Supercomputers in Stellar
Dynamics}, eds. P. Hut and S. McMillan (New York: Springer),
pp. 212-218, 1986.}

\note{5}{D. Sugimoto \& E. Bettwieser, Post-Collapse Evolution
of Globular Clusters, Mon. Not. R. astr. Soc., 204,
pp. 19p-22p, 1983.; E. Bettwieser, E. \& D. Sugimoto,
Post-Collapse Evolution and Gravothermal Oscillations of
Globular Clusters, Mon. Not. R.  astr. Soc., 208, pp. 493-509,
1984.}

\note{6}{J. Goodman, On Gravothermal Oscillations, Astroph. J.,
313, pp. 576-595, 1987.} 

\note{7}{H. Cohn, P. Hut \& M. Wise, Gravothermal Oscillations
after Core Collapse in Globular Cluster Evolution,
Astrophys. J., 342: pp. 814-822, 1989.  J. L. Breeden,
H. N. Cohn \& P. Hut, The Onset of Gravothermal Oscillations in
Globular Cluster Evolution, Astrophys. J., 421: pp. 195-205, 1994.
R. Spurzem \&, P. D. Louis, in {\it Structure and Dynamics of
Globular Clusters}, eds. S. G. Djorgovski and G. Meylan, ASP
Conf. Ser. Vol. 50 (San Francisco: Astron. Soc. Pacific), 1993,
p.135.}

\note{8}{J.L. Breeden \& H.N. Cohn, Chaos in Core Oscillations
of Globular Clusters, Astrophys. J., 448, pp. 672-682, 1995.}

\note{9}{J. Makino, Gravothermal Oscillations, in Dynamical
Evolution of Star Clusters, I.A.U. Symp. 174, eds. P. Hut and
J. Makino (Dordrecht: Kluwer), pp. 151-160, 1996.}

\note{10}{J. Makino \& M. Taiji, Computational Astrophysics on a
Special-Purpose Machine, Computers in Physics, 10, pp. 352-358,
1996.}

\note{11}{Guhathakurta, P., Yanny, B., Schneider, D.P. \&
Bahcall, J.N., Globular Cluster Photometry with the Hubble Space
Telescope V.  WFPC2 Study of M15's Central Density Cusp,
Astron. J., 111, pp. 267-282, 1996.}

\note{12}{P. Hut, J. Makino \& S. McMillan, Modelling Globular
Cluster Evolution, Nature, 336: pp. 31-35, 1988.}

\note{13}{S. F. Portegies Zwart, P. Hut \& F. Verbunt, Star
Cluster Ecology I, Astron. \& Astrophys., in press, 1997;
S. F. Portegies Zwart, P. Hut, S. L. W. McMillan \& F. Verbunt,
Star Cluster Ecology II, Astron. \& Astrophys., in press, 1997;
S. F. Portegies Zwart, P. Hut, J. Makino, S. L. W. McMillan \&
F. Verbunt, Star Cluster Ecology III, Astron. \& Astrophys., in
preparation, 1997}

\note{14}{S. J. Aarseth, Star Cluster Simulations on HARP, in
Dynamical Evolution of Star Clusters, I.A.U. Symp. 174,
eds. P. Hut and J. Makino (Dordrecht: Kluwer), pp. 161-170,
1996.}

  \bye